\documentclass[journal]{IEEEtran}

\usepackage[usenames, dvipsnames]{color}
\usepackage{epstopdf}
\usepackage{graphicx}
\usepackage{amsmath}
\usepackage{amssymb}
\usepackage{epstopdf}
\usepackage{algorithmic}
\usepackage{subfigure}
\usepackage{fancybox}
\usepackage{multicol}
\usepackage{verbatim}
\usepackage{caption}
\usepackage{bm}
\usepackage[pdfborder={1 1 1},pdfborderstyle={/S/U/W 1}]{hyperref}
\usepackage[ruled,vlined]{algorithm2e}
\usepackage{marginnote,lipsum}
\begin{document}

\title{A Generalized Structured Low-Rank Matrix Completion Algorithm for MR Image Recovery}

\author{Yue~Hu,~\IEEEmembership{Member,~IEEE,}
        Xiaohan~Liu,~\IEEEmembership{Student~Member,~IEEE,}
        and~Mathews~Jacob,~\IEEEmembership{Senior~Member,~IEEE}
\thanks{This work was supported by grants Natural Science Foundation of China (NSFC) 61501146, Natural Science Foundation of Heilongjiang province F2016018, and NIH 1R01EB019961-01A1.}
\thanks{Y. Hu and X. Liu are with the School
of Electronics and Information Engineering, Harbin Institute of Technology, Harbin,
China 150001 (e-mail: huyue@hit.edu.cn).}
\thanks{M. Jacob is with the Department of Electrical and Computer Engineering, University of Iowa, IA 52246, USA (email: mathews-jacob@uiowa.edu).}
}


\maketitle

\begin{abstract}
Recent theory of mapping an image into a structured low-rank Toeplitz or Hankel matrix has become an effective method to restore images. In this paper, we introduce a generalized structured low-rank algorithm to recover images from their undersampled Fourier coefficients using infimal convolution regularizations. The image is modeled as the superposition of a piecewise constant component and a piecewise linear component. The Fourier coefficients of each component satisfy an annihilation relation, which results in a structured Toeplitz matrix, respectively. We exploit the low-rank property of the matrices to formulate a combined regularized
optimization problem. In order to solve the problem efficiently and to avoid the high memory demand resulting from the large-scale Toeplitz matrices, we introduce a fast and memory efficient algorithm based on the half-circulant approximation of the Toeplitz matrix. We demonstrate our algorithm in the context of single and multi-channel MR images recovery. Numerical experiments indicate that the proposed algorithm provides improved recovery performance over the state-of-the-art approaches.
\end{abstract}


\IEEEpeerreviewmaketitle

\section{Introduction}

The recovery of images from their limited and noisy measurements is an important problem in a wide range of biomedical imaging applications, including microscopy \cite{dey2006richardson}, magnetic resonance imaging (MRI) \cite{lustig2007sparse}, and computed tomography \cite{sidky2008image}. The common method is to formulate the image reconstruction as an optimization problem, where the criterion is a linear combination of data consistency error and a regularization penalty. The regularization penalties are usually chosen to exploit the smoothness or the sparsity priors in the discrete image domain. For example, compressed sensing methods are capable of recovering the original MR images from their partial $k$ space measurements \cite{lustig2007sparse} using the $L_1$ norm in the total variation (TV) or wavelet domain. The reconstruction performance is determined by the effectiveness of the regularization. In order to improve the quality of the reconstructed images, several extensions and generalizations of TV are also proposed, such as total generalized variation (TGV) \cite{knoll2011second,knoll2012parallel}, Hessian-based norm regularization \cite{lefkimmiatis2012hessian}, and higher degree total variation (HDTV) \cite{hu2012higher,hu2014generalized}. All of these regularization penalties are formulated in the discrete domain, and hence suffer from discretization errors as well as lack of rotation invariance.

Recently, a new family of reconstruction methods, which are based on the low-rank property of structured Hankel or Toeplitz matrices built from the Fourier coefficients of the image, have been introduced as powerful continuous domain alternatives to the above discrete domain penalties \cite{jin2016general,haldar2014low,ongie2016off,pan2014sampling,ongie2015recovery}. Since these methods minimize discretization errors and rotation dependence, they provide improved reconstruction. These algorithms can be viewed as the multidimensional extensions of the finite-rate-of-innovation (FRI) framework \cite{vetterli2002sampling,maravic2005sampling}. All of these methods exploit the ``annihilation property", which implies that image derivatives can be annihilated by multiplication with a bandlimited polynomial function in the spatial domain; this image domain relation translates to a convolutional annihilation relationship in the Fourier domain. Since the locations of the discontinuities are not isolated in the multidimensional setting, the theoretical tools used to show perfect recovery are very different from the 1-D FRI setting \cite{ongie2016off,ongie2017convex}. The convolution relations are compactly represented as a multiplication between a block Hankel structured matrix and the Fourier coefficients of the filter. It has been shown that the above structured matrix is low-rank, which allows the recovery of the unknown matrix entries using structured low-rank matrix completion. Empirical results show improved performance over classical total variation methods \cite{ongie2017fast,ongie2017convex, jin2016general,jin2015novel}. Haldar proposed a Hankel structured low-rank matrix algorithm (LORAKS) for the reconstructions of single coil MR images \cite{haldar2014low} with the assumption that the image has limited spatial support and smooth phase. The effectiveness of the algorithm was also investigated in parallel MRI \cite{shin2014calibrationless,haldar2016p,kim2017loraks}.

In this paper, we extend the structured low-rank framework to recover the sum of two piecewise smooth functions from their sparse measurements. This work is inspired by the infimal convolution framework \cite{chambolle1997image}, where the sum of a piecewise constant and a piecewise linear function was recovered; the infimal convolution of functions with first and second order derivatives were considered as penalties in \cite{chambolle1997image}. The algorithm was then applied in a general discrete setting for image denoising \cite{setzer2011infimal} to obtain improved performance over standard TV. The extension of TV using infimal convolution (ICTV) was applied in video and image reconstruction in \cite{holler2014infimal}. The infimal convolution of TGV (ICTGV) was proposed in the context of dynamic MRI reconstruction by balancing the weights for the spatial and temporal regularizations \cite{schloegl2017infimal}. In \cite{rasch2017joint} and \cite{rasch2018dynamic}, the infimal convolution of two total variation Bregman distances are applied to exploit the structural information in the reconstruction of dynamic MR datasets and the joint reconstruction of PET-MRI, respectively. In \cite{otazo2015low}, the authors adopted the robust PCA method \cite{candes2011robust} into dynamic MRI, where the dataset is decomposed into low-rank and sparse component (L+S). In \cite{velikina2015reconstruction}, instead of imposing low-rank assumptions, the k-t PCA method was improved using model consistency constraints (MOCCO) to obtain temporal basis functions from low resolution dynamic MR data. In this paper, we propose to model the image as the combination of a piecewise constant component and a piecewise linear component. For the piecewise constant component, 
the Fourier coefficients of the gradient of the component satisfy the annihilation relation. We thus build a structured Toeplitz matrix, which can be proved to be low-rank. Similarly, we can obtain a structured low-rank Toeplitz matrix from the Fourier coefficients of the second order partial derivatives of the piecewise linear component. By introducing the generalized structured low-rank method, the image can be automatically separated into components where either the strong edges and feature details or the smooth regions of the image can be accurately recovered. Thus, the optimal balance can be obtained between the first order and higher order penalties.

Since the proposed method involves the recovery of large-scale first and second order derivatives lifted Toeplitz matrices, the implementation of the method is associated with high computational complexity and memory demand. In order to solve the corresponding optimization problem efficiently, we introduce an algorithm based on the half-circulant approximation of Toeplitz matrices, which is a generalization of the Generic Iteratively Reweighted Annihilating Filter (GIRAF) algorithm proposed in \cite{ongie2017fast,ongie2016fast}.
This algorithm alternates between the estimation of the annihilation filter of the image, and the computation of the image annihilated by the filter in a least squares formulation. By replacing the linear convolution by circular convolution, the algorithm can be implemented efficiently using Fast Fourier Transforms, which significantly reduces the computational complexity and the memory demand. We investigate the performance of the algorithm in the context of compressed sensing MR images reconstruction. Experiments show that the proposed method is capable of providing more accurate recovery results than the state-of-the-art algorithms. The preliminary version of the work was accepted as a conference paper. Compared to the work \cite{hu2018adaptive}, the theoretical and algorithmic frameworks are further developed here. We have significantly more validation in the current version, in addition to the generalization of the method to parallel MRI setting.


\section{Generalized Structure Low-Rank Matrix Recovery}
\subsection{Image recovery model}
We consider the recovery of a discrete 2-D image $\boldsymbol{ \rho}\in\mathbb C^N$ from its noisy and degraded measurements $\mathbf b \in \mathbb C^M$. We model the measurements as $\mathbf b = {\cal A}(\boldsymbol\rho)+\mathbf n$, where ${\cal A} \in\mathbb C^{M\times N}$ is a linear degradation operator which maps $\boldsymbol\rho $ to $\mathbf b $, and $\mathbf n \in \mathbb C^M$ is assumed to be the Gaussian distributed white noise. Since the recovery of $\boldsymbol\rho$ from the measurements $\mathbf b$ is ill-posed in many practical cases, the general approach is to pose the recovery as a regularized optimization problem, i.e.,
\begin{equation}
\boldsymbol \rho = \arg\min_{\boldsymbol \rho}\|{\cal A}(\boldsymbol \rho)-\mathbf b\|^2+\lambda{\cal J}(\boldsymbol \rho)
\end{equation}
where $\|{\cal A}(\boldsymbol \rho)-\mathbf b\|^2$ is the data consistency term, $\lambda$ is the balancing parameter, and ${\cal J}(\boldsymbol \rho)$ is the regularization term, which determines the quality of the recovered image. Common choices for the regularization term include total variation, wavelet, and their combinations. Researchers have also proposed the extensions of total variations \cite{knoll2011second,lefkimmiatis2012hessian,hu2014generalized} to improve the performance.

\subsection{Structured low-rank matrix completion}\label{rho1_rho2}
Consider the general model for a 2-D piecewise smooth image $\rho(\mathbf r)$ at the spatial location $\mathbf r=(x,y)\in\mathbb Z^2$:
\begin{equation}
\label{eq:piecesmooth}
\rho(\mathbf r)=\sum_{i=1}^{N}g_i(\mathbf r)\chi_{\Omega_i}(\mathbf r)
\end{equation}
where $\chi_{\Omega_i}$ is a characteristic function of the set $\Omega_i$ and the functions $g_i(\mathbf r)$ are smooth polynomial functions which vanish with a collection of differential operators $\mathbf D=\{D_1,...,D_N\}$ within the region $\Omega_i$. It is proved that under certain assumptions on the edge set $\partial \Omega=\bigcup _{i=1}^N\partial \Omega_i$, the Fourier transform of derivatives of $\rho(\mathbf r)$ satisfies an annihilation property \cite{ongie2016off}. We assume that a bandlimited trigonometric polynomial function $\mu(\mathbf r)$ vanishes on the edge set of the image:
\begin{equation}\label{eq:mu}
\mu(\mathbf r)=\sum_{\mathbf k \in \Delta_1}c[\mathbf k]e^{j2\pi\langle\mathbf k,\mathbf r\rangle}
\end{equation}
where $c[\mathbf k]$ denotes the Fourier coefficients of $\mu$ and $\Delta_1$ is any finite set of $\mathbb Z^2$.
According to \cite{ongie2015recovery}, the family of functions in (\ref{eq:piecesmooth}) is a general form including many common image models by choosing different set of differential operators $\mathbf D$.

\begin{enumerate}
\item{Piecewise constant images: }Assume $\rho_1(\mathbf r)$ is a piecewise constant image function, thus the first order partial derivative of the image $\mathbf D_1\rho_1=\bm{\nabla} \rho_1=(\partial_x \rho_1,\partial_y \rho_1)$ is annihilated by multiplication with $\mu_1$ in the spatial domain:
    \begin{equation}
    \mu_1\bm{\nabla}\rho_1=0
    \end{equation}
    The multiplication in spatial domain translates to the convolution in Fourier domain, which is expressed as:
    \begin{equation}
    \sum_{\mathbf k\in\Delta_1}\widehat{\bm{\nabla}\rho_1}[\bm \ell-\mathbf k]c_1[\mathbf k]=\mathbf 0,\;\forall\bm\ell\in\mathbb Z^2
    \end{equation}
    where $\widehat{\bm\nabla \rho_1}[\mathbf k]=j2\pi(k_x\hat\rho_1[\mathbf k],k_y\hat\rho_1[\mathbf k])$ for $\mathbf k=(k_x,k_y)$. Thus the annihilation property can be formulated as a matrix multiplication:
    \begin{equation}
    {\cal T}_1(\hat \rho_1)\mathbf c=\left[\begin{array}{c}
    {\cal T}_x(\hat\rho)\\{\cal T}_y(\hat \rho)\end{array}\right]\mathbf c_1=\mathbf 0
    \end{equation}
    where ${\cal T}_1(\hat\rho_1)$ is a Toeplitz matrix built from the entries of $\hat \rho_1$, the Fourier coefficients of $\rho_1$.
    Specifically, ${\cal T}_x(\hat\rho_1)$, ${{\cal T}_y(\hat\rho_1)}$ are matrices corresponding to the discrete 2-D convolution of $k_x\hat\rho_1[\mathbf k]$ and $k_y\hat\rho_1[\mathbf k]$ for $(k_x,k_y)\in\Gamma$, omitting the irrelevant factor $j2\pi$. Here $\mathbf c_1$ is the vectorized version of the filter $c_1[\mathbf k]$.
    Note that $\mathbf c$ is supported in $\Delta_1$. Thus, we can obtain that:
    \begin{equation}
    \hat\rho_1[\mathbf k]*d[\mathbf k]=0, \;\mathbf k\in \Gamma
    \end{equation}
    Here $d[\mathbf k]=c_1[\mathbf k]*h[\mathbf k]$, where $h[\mathbf k]$ is any FIR filter. Note that $\Delta_1$ is smaller than $\Gamma$, the support of $d$. Thus, if we take a larger filter size than the minimal filter $c_1[\mathbf k]$, the annihilation matrix will have a larger null space. Therefore, ${\cal T}_1(\hat\rho_1)$ is a low-rank matrix. The method corresponding to this case is referred to as the first order structured low-rank algorithm (first order SLA) for simplicity.

 \item{Piecewise linear images:}
 Assume $\rho_2(\mathbf r)$ is a piecewise linear image function, it can be proved that the second order partial derivatives of the image $\mathbf D_2\rho_2=(\partial^2_{xx}\rho_2,\partial^2_{xy}\rho_2,\partial^2_{yy}\rho_2)$ satisfy the annihilation property \cite{ongie2015recovery}:
 \begin{equation}
 \mu_2^2 \mathbf D_2\rho_2=0
 \end{equation}
 Thus, the Fourier transform of $\mathbf D_2\rho_2$ is annihilated by convolution with the Fourier coefficients $c_2[\mathbf k],\mathbf k\in \Delta_2$ of $\mu_2^2$:
 \begin{equation}
 \sum_{\mathbf k\in\Delta_2}\widehat{\mathbf D_2\rho_2}[\bm \ell-\mathbf k]c_2[\mathbf k]=\mathbf 0,\;\forall\bm\ell\in\mathbb Z^2
 \end{equation}
 where $\widehat{\mathbf D_2\rho_2}[\mathbf k]=(j2\pi)^2(k_x^2\hat\rho_2[\mathbf k],k_xk_y\hat\rho_2[\mathbf k],k_y^2\hat\rho_2[\mathbf k])$ for $\mathbf k=(k_x,k_y)$. Similarly, the annihilation relation can be expressed as:
 \begin{equation}
 {\cal T}_2(\hat\rho_2)\mathbf d=\left[\begin{array}{c}
    {\cal T}_{xx}(\hat\rho_2)\\{\cal T}_{xy}(\hat \rho_2)\\{\cal T}_{yy}(\hat\rho_2)\end{array}\right]\mathbf c_2=\mathbf 0
 \end{equation}
 where ${\cal T}_2(\hat\rho_2)$ is a Toeplitz matrix. ${\cal T}_{xx}(\hat\rho_2)$, ${{\cal T}_{xy}(\hat\rho_2)}$, and ${{\cal T}_{yy}(\hat\rho_2)}$ are matrices corresponding to the discrete convolution of $k_x^2\hat\rho_2[\mathbf k]$, $k_xk_y\hat\rho_2[\mathbf k]$, and $k_y^2\hat\rho_2[\mathbf k]$, omitting the insignificant factor, and $\mathbf c_2$ is the vectorized version of $d[\mathbf k]$. Similar to the piecewise constant case, the Toeplitz matrix ${\cal T}_2(\hat\rho_2)$ is also a low-rank matrix. The method exploiting the low-rank property of ${\cal T}_2(\hat\rho_2)$ is referred to as the second order structured low-rank algorithm (second order SLA).

\end{enumerate}
\subsection{Generalized structured low-rank image recovery (GSLR)}

We assume that a 2-D image $\boldsymbol\rho$ is a piecewise smooth function, which can be decomposed into two components $\boldsymbol\rho=\boldsymbol\rho_1+\boldsymbol\rho_2$, such that $\boldsymbol\rho_1$ represents the piecewise constant component of $\boldsymbol\rho$, while $\boldsymbol\rho_2$ represents the piecewise linear component of $\boldsymbol\rho$.
We assume that the gradient of $\boldsymbol \rho_1$ and the second derivative of $\boldsymbol\rho_2$ vanish on the zero sets of $\mu_1$ and $\mu_2$, respectively. This relation transforms to a convolution relation between the weighted Fourier coefficients of $\boldsymbol\rho_1$ and $\boldsymbol\rho_2$ with $c_1$ and $c_2$, respectively, based on the analysis in Section. \ref{rho1_rho2}.
Inspired by the concept of infimal convolution, we consider the framework of a combined regularization procedure, where we formulate the reconstruction of the Fourier data $\hat{\boldsymbol\rho}$ from the undersampled measurements $\mathbf b$ as follows:
\begin{align}\label{eq:rankprob}
\min_{\hat{\boldsymbol\rho}_1+\hat{\boldsymbol\rho}_2=\hat{\boldsymbol\rho}}\;\;\mbox{rank}[{\cal T}_1(\hat{\boldsymbol\rho}_1)]+\mbox{rank}[{\cal T}_2(\hat{\boldsymbol\rho}_2)]\;\;\nonumber\\
\mbox{such}\;\mbox{that}\;\mathbf b={\cal A}(\hat{\boldsymbol\rho})+\mathbf n
\end{align}
Since the above problem is NP hard, we choose the Schatten $p\;(0\leq p < 1)$ norm as the relaxation function, which makes (\ref{eq:rankprob}) as the following optimization problem:
\begin{align}\label{eq:optprob}
\{\hat{\boldsymbol\rho_1}^\star,\hat{\boldsymbol\rho_2}^\star\}=\arg\min_{\hat{\boldsymbol\rho_1},\hat{\boldsymbol\rho_2}}
\lambda_1\|{\cal T}_1(\hat{\boldsymbol\rho_1})\|_p+\lambda_2\|{\cal T}_2(\hat{\boldsymbol\rho_2})\|_p \nonumber\\
+\|{\cal A}(\hat{\boldsymbol\rho_1}+\hat{\boldsymbol\rho_2})-\mathbf b \|^2
\end{align}
where ${\cal T}_1(\hat{\boldsymbol\rho_1})$ and ${\cal T}_2(\hat{\boldsymbol\rho_2})$ are the structured Toeplitz matrices in the first and second order partial derivatives weighted lifted domain, respectively. $\lambda_1$ and $\lambda_2$ are regularization parameters which balance the data consistency and the degree to which ${\cal T}_1(\hat{\boldsymbol\rho_1})$ and ${\cal T}_2(\hat{\boldsymbol\rho_2})$ are low-rank. $\|\cdot\|_p$ is the Schatten $p$ norm ($0\leq p<1$), defined for an arbitrary matrix $\mathbf X$ as:
 \begin{equation}
 \|\mathbf X\|_p = \frac{1}{p}\mbox{Tr}[(\mathbf X^*\mathbf X)^{\frac{p}{2}}]=\frac{1}{p}\mbox{Tr}[(\mathbf X\mathbf X^*)^{\frac{p}{2}}]=\frac{1}{p}\sum\limits_i\sigma_i^p
 \end{equation}
  where $\sigma_i$ are the singular values of $\mathbf X$. Note that when $p\rightarrow 1$, $\|\mathbf X\|_p$ is the nuclear norm; when $p\rightarrow 0$, $\|\mathbf X\|_p=\sum\limits_i \log \sigma_i$. The penalty $\|\mathbf X\|_p$ is convex for $p=1$, and non-convex for $0\leq p<1$.

 \section{Optimization Algorithm}

\label{sec:pagestyle}

We apply the iterative reweighted least squares (IRLS) algorithm to solve the optimization problem (\ref{eq:optprob}). Based on the equation $\|\mathbf X\|_p=\|\mathbf X\mathbf H^{\frac{1}{2}}\|_F^2$, where $\mathbf H=(\mathbf X^*\mathbf X)^{\frac{p}{2}-1}$, let $\mathbf X={\cal T}_i(\hat{\boldsymbol\rho}_i)$ ($i=1,2$), (\ref{eq:optprob}) becomes:
\begin{align}\label{eq:optprob1}
\{\hat{\boldsymbol\rho}_1^\star,\hat{\boldsymbol\rho}_2^\star\}=&\arg\min_{\hat{\boldsymbol\rho}_1,\hat{\boldsymbol\rho}_2}
\lambda_1\|{\cal T}_1(\hat{\boldsymbol\rho}_1)\mathbf H_1^{\frac{1}{2}}\|^2_F+\lambda_2\|{\cal T}_2(\hat{\boldsymbol\rho}_2)\mathbf H_2^{\frac{1}{2}}\|^2_F\nonumber\\
&+\|{\cal A}(\hat{\boldsymbol\rho}_1+\hat{\boldsymbol\rho}_2)-\mathbf b \|^2
\end{align}
In order to solve (\ref{eq:optprob1}), we can use an alternating minimization scheme, which alternates between the following subproblems: updating the weight matrices $\mathbf H_1$ and $\mathbf H_2$, and solving a weighted least squares problem. Specifically, at $n$th iteration, we compute:
\begin{equation}\label{eq:H1}
\mathbf H_{1}^{(n)}=[\underbrace{{\cal T}_1(\hat{\boldsymbol\rho}_{1}^{(n)})^*{\cal T}_1(\hat{\boldsymbol\rho}_{1}^{(n)})}_{\mathbf G_1}+\epsilon_n\mathbf I]^{\frac{p}{2}-1}
\end{equation}
\begin{equation}\label{eq:H2}
\mathbf H_{2}^{(n)}=[\underbrace{{\cal T}_2(\hat{\boldsymbol\rho}_{2}^{(n)})^*{\cal T}_2(\hat{\boldsymbol\rho}_{2}^{(n)})}_{\mathbf G_2}+\epsilon_n\mathbf I]^{\frac{p}{2}-1}
\end{equation}

\begin{align}\label{eq:leastsquare}
&\{\hat{\boldsymbol\rho}_{1}^{(n)},\hat{\boldsymbol\rho}_{2}^{(n)}\}=\arg\min_{\hat{\boldsymbol\rho}_1,\hat{\boldsymbol\rho}_2}
\|{\cal A}(\hat{\boldsymbol\rho}_1+\hat{\boldsymbol\rho}_2)-\mathbf b \|^2\nonumber\\
&
+\lambda_1\|{\cal T}_1(\hat{\boldsymbol\rho}_1) (\mathbf H_{1}^{(n)})^{\frac{1}{2}}\|^2_F
+\lambda_2\|{\cal T}_2(\hat{\boldsymbol\rho}_2)(\mathbf H_{2}^{(n)})^{\frac{1}{2}}\|^2_F
\end{align}

where $\epsilon_n\rightarrow0$ is a small factor used to stabilize the inverse. We now show how to efficiently solve the subproblems.
\subsection{Update of least squares}

First, let ${\mathbf H}_1^{\frac{1}{2}}=[\mathbf h_1^{(1)},\ldots,\mathbf h_1^{(L)}]$, ${\mathbf H}_2^{\frac{1}{2}}=[\mathbf h_2^{(1)},\ldots,\mathbf h_2^{(M)}]$, we rewrite the least squares problem (\ref{eq:leastsquare}) as follows:
\begin{align}\label{eq:leastsquare1}
&\{\hat{\boldsymbol\rho}_{1}^{(n)},\hat{\boldsymbol\rho}_{2}^{(n)}\}=\arg\min_{\hat{\boldsymbol\rho}_1,\hat{\boldsymbol\rho}_2}
\|{\cal A}(\hat{\boldsymbol\rho}_1+\hat{\boldsymbol\rho}_2)-\mathbf b \|^2 \nonumber\\
&+\lambda_1\sum_{l=1}^L\|{\cal T}_1(\hat{\boldsymbol\rho}_1)\mathbf h_{1}^{(l)}\|^2_F
+\lambda_2\sum_{m=1}^M\|{\cal T}_2(\hat{\boldsymbol\rho}_2)\mathbf h_{2}^{(m)}\|^2_F
\end{align}
We now focus on the update of $\hat{\boldsymbol\rho_1}$. The update of  $\hat{\boldsymbol\rho_2}$ can be derived likewise. From the structure property of ${\cal T}_1(\hat{\boldsymbol\rho_1})$ and the convolution relationship, we can obtain:
\begin{align}\label{eq:1stsimp}
{\cal T}_1(\hat{\boldsymbol\rho_1})\mathbf h_1^{(l)}&={\cal P}_{\Gamma_1}(\mathbf M_1\hat{\boldsymbol\rho_1}*\mathbf h^{(l)}_1)={\cal P}_{\Gamma_1}(\mathbf h^{(l)}_1*\mathbf M_1\hat{\boldsymbol\rho_1})\nonumber\\&=\mathbf P_1 \mathbf C_1^{(l)}\mathbf M_1\hat{\boldsymbol\rho_1},l=1,...,L
\end{align}
where $\mathbf C_1^{(l)}$ denotes the linear convolution by $\mathbf h_1^{(l)}$, ${\cal P}_{\Gamma_1}$ is the projection of the convolution to a finite set $\Gamma_1$ of the valid $k$ space index, which is expressed by the matrix $\mathbf P_1$. $\mathbf M_1$ is the linear transformation in $k$ space, which denotes the multiplication with the first order Fourier derivatives $j2\pi k_x$ and $j2\pi k_y$, referred to as the gradient weighted lifting case.
We can approximate $\mathbf C_1^{(l)}$ by a circular convolution by $\mathbf h_1^{(l)}$ on a sufficiently large convolution grid. Then, we can obtain $\mathbf C_1^{(l)}=\mathbf F\mathbf S_1^{(l)}\mathbf F^*$, where $\mathbf F$ is the 2-D DFT and $\mathbf S_1^{(l)}$ is a diagonal matrix representing multiplication with the inverse DFT of $\mathbf h_1^{(l)}$.
Assuming $\mathbf P_1^*\mathbf P_1\approx \mathbf I$, we can thus rewrite the second term in (\ref{eq:leastsquare1}) as:
\begin{align}
\lambda_1\sum_{l=1}^L\|\mathbf P_1\mathbf C_1^{(l)}\mathbf M_1\hat{\boldsymbol\rho_1}\|^2&=\lambda_1\hat{\boldsymbol\rho_1}^*\mathbf M_1^*\mathbf F\sum_{l=1}^L\mathbf S_1^{(l)*}\mathbf S_1^{(l)}\mathbf F^*\mathbf M_1\hat{\boldsymbol\rho_1}\nonumber\\
&=\lambda_1\|\mathbf S_1^{\frac{1}{2}}\mathbf F^*\mathbf M_1\hat{\boldsymbol\rho_1}\|^2
\end{align}
where $\mathbf S_1$ is a diagonal matrix with entries $\sum_{l=1}^N|\mu_l(\mathbf r)|^2$, and $\mu_l(\mathbf r)$ is the trigonometric polynomial of inverse Fourier transform of $\mathbf h_1^{(l)}$. $\mathbf S_1$ is specified as
\begin{equation}\label{eq:s1}
\mathbf S_1=\sum\limits_{l=1}^L \mathbf S_1^{(l)*}\mathbf S_1^{(l)}=\mbox{diag}(\mathbf F^*\mathbf P_1^*\mathbf h_1)
\end{equation}

Similarly, the third term in (\ref{eq:leastsquare1}) can be rewritten as $\lambda_2\|\mathbf S_2^{\frac{1}{2}}\mathbf F^*\mathbf M_2\hat{\boldsymbol\rho_2}\|^2$, where $\mathbf S_2$ is given by
\begin{equation}\label{eq:s2}
\mathbf S_2=\sum\limits_{m=1}^M \mathbf S_2^{(m)*}\mathbf S_2^{(m)}=\mbox{diag}(\mathbf F^*\mathbf P_2^*\mathbf h_2)
\end{equation}

 Therefore, we can reformulate the optimization problem (\ref{eq:leastsquare1}) as:
\begin{align}\label{eq:newprob}
\min_{\hat{\boldsymbol\rho}_1,\hat{\boldsymbol\rho}_2}
&\|{\cal A}(\hat{\boldsymbol\rho}_1+\hat{\boldsymbol\rho}_2)-\mathbf b \|^2
+\lambda_1\|\mathbf S_1^{\frac{1}{2}}\mathbf y_1\|^2_F+\lambda_2\|\mathbf S_2^{\frac{1}{2}}\mathbf y_2\|^2_F
 \nonumber\\
&\mbox{s.t.}\;\mathbf F\mathbf y_1=\mathbf M_1\hat{\boldsymbol\rho}_1,\;\mathbf F\mathbf y_2=\mathbf M_2\hat{\boldsymbol\rho}_2
\end{align}

The above constrained problem can be efficiently solved using the alternating directions method of multipliers (ADMM) algorithm \cite{esser2009applications}, which yields to solving the following subproblems:
\begin{equation}\label{y1}
\mathbf y_1^{(n)}=\min_{\mathbf y_1}\|\mathbf S_1^\frac{1}{2}\mathbf y_1\Vert_2^2+\gamma_1\|\mathbf q_1^{(n-1)}+\mathbf F^*\mathbf M_1{\hat{\boldsymbol\rho_1}}^{(n-1)}-{\mathbf y}_1\Vert_2^2
\end{equation}

\begin{equation}\label{y2}
\mathbf y_2^{(n)}=\min_{\mathbf y_2}\|\mathbf S_2^\frac{1}{2}\mathbf y_2\Vert_2^2+\gamma_2\|\mathbf q_2^{(n-1)}+\mathbf F^*\mathbf M_2{\hat{\boldsymbol\rho_2}}^{(n-1)}-\mathbf y_2\Vert_2^2
\end{equation}

\begin{small}
\begin{equation}\label{rho1}
\hat{\boldsymbol\rho}_1^{(n)}=\min_{\hat{\boldsymbol\rho}_1}\|{\cal A}(\hat{\boldsymbol\rho}_1+\hat{\boldsymbol\rho}_2)-\mathbf b\Vert_2^2+\gamma_1 \lambda_1\|\mathbf q_1^{(n-1)}+\mathbf F^*\mathbf M_1\hat{\boldsymbol\rho}_1-\mathbf y_1^{(n)}\Vert_2^2
\end{equation}
\end{small}

\begin{small}
\begin{equation}\label{rho2}
\hat{\boldsymbol\rho}_2^{(n)}=\min_{\hat\rho_2}\|{\cal A}(\hat{\boldsymbol\rho}_1+\hat{\boldsymbol\rho}_2)-\mathbf b\Vert_2^2
+\gamma_2 \lambda_2\|\mathbf q_2^{(n-1)}+\mathbf F^*\mathbf M_2\hat{\boldsymbol\rho}_2-\mathbf y_2^{(n)}\Vert_2^2
\end{equation}
\end{small}

\begin{equation}\label{eq:qi}
\mathbf q_i^{(n)}=\mathbf q_i^{(n-1)}+\mathbf F^*\mathbf M_{i}\hat{\boldsymbol\rho}_{i}^{(n)}-\mathbf y_{i}^{(n)}\quad i=1,2
\end{equation}

where $\mathbf q_{i}$ ($i=1,2$) represent the vectors of Lagrange multipliers, and $\gamma_{i}$ ($i=1,2$)  are fixed parameters tuned to improve the conditioning of the subproblems. Subproblems (\ref{y1}) to (\ref{rho2}) are quadratic and thus can be solved easily as follows:

\begin{equation}
\mathbf y_1^{(n)}=(\mathbf S_1+\gamma_1\mathbf I)^{-1}[\gamma_1(\mathbf q_1^{(n-1)}+\mathbf F^*\mathbf M_1\hat{\boldsymbol\rho}_1^{(n-1)})]
\end{equation}
\begin{equation}
\mathbf y_2^{(n)}=(\mathbf S_2+\gamma_2\mathbf I)^{-1}[\gamma_2(\mathbf q_2^{(n-1)}+\mathbf F^*\mathbf M_2 \hat{\boldsymbol\rho}_2^{(n-1)})]
\end{equation}
\begin{equation}
\begin{split}
\hat{\boldsymbol\rho}_1^{(n)}=({\cal A}^*{\cal A}+\gamma_1\lambda_1\mathbf M_1^*\mathbf M_1)^{-1}[&\gamma_1\lambda_1(\mathbf M_1^*\mathbf F)(\mathbf y_1^{(n)}-\mathbf q_1^{(n-1)})\\
&+{\cal A}^*\mathbf b-{\cal A}^*{\cal A}\hat{\boldsymbol\rho}_2^{(n-1)}]
\end{split}
\end{equation}
\begin{equation}
\begin{split}
\hat{\boldsymbol\rho}_2^{(n)}=({\cal A}^*{\cal A}+\gamma_2\lambda_2\mathbf M_2^*\mathbf M_2)^{-1}[&\gamma_2\lambda_2(\mathbf M_2^*\mathbf F)(\mathbf y_2^{(n)}-\mathbf q_2^{(n-1)})\\
&+{\cal A}^*\mathbf b-{\cal A}^*{\cal A}\hat{\boldsymbol\rho}_1^{(n-1)}]
\end{split}
\end{equation}

\subsection{Update of weight matrices}

We now show how to update the weight matrices $\mathbf H_1$ and $\mathbf H_2$ in (\ref{eq:H1}) and (\ref{eq:H2})
efficiently based on the GIRAF method \cite{ongie2016off}.
In order to obtain the weight matrices $\mathbf H_1$ and $\mathbf H_2$, we first compute the Gram matrix $\mathbf G_1$ and $\mathbf G_2$ as:
\begin{align}
\label{eq:grammtx1}
\mathbf G_1={\cal T}_1(\hat{\boldsymbol\rho}_{1})^*{\cal T}_1(\hat{\boldsymbol\rho}_{1})\\
\label{eq:grammtx2}
\mathbf G_2={\cal T}_2(\hat{\boldsymbol\rho}_{2})^*{\cal T}_2(\hat{\boldsymbol\rho}_{2})
\end{align}

Let $(\mathbf V_1,\bm{\Lambda}_1)$ denote the eigen-decomposition of $\mathbf G_1$, where $\mathbf V_1$ is the orthogonal basis of eigenvectors $\mathbf v_1^{(l)}$, and $\bm{\Lambda_1}$ is the diagonal matrix of eigenvalues $\lambda_{p_1}^{(l)}$, which satisfy $\mathbf G_1=\mathbf V_1 \bm{\Lambda}_1 \mathbf V_1^*$. Then we can rewrite the weight matrix $\mathbf H_1$ as:
\begin{equation}
\mathbf H_1=[\mathbf V_1(\bm{\Lambda}_1+\epsilon\mathbf I)\mathbf V_1^*]^{\frac{p}{2}-1}=\mathbf V_1(\bm{\Lambda}_1+\epsilon\mathbf I)^{\frac{p}{2}-1}\mathbf V_1^*
\end{equation}
Thus, one choice of the matrix square root $\mathbf H_1^{\frac{1}{2}}$ is
\begin{align}\label{eq:h1}
\mathbf H_1^{\frac{1}{2}}&=(\bm\Lambda_1+\epsilon\mathbf I)^{\frac{p}{4}-\frac{1}{2}}\mathbf V_1^*\nonumber\\
&=[(\lambda_{p_1}^{(1)}+\epsilon)^{\frac{p}{4}-\frac{1}{2}} (\mathbf v_1^{(1)})^*,...,(\lambda_{p_1}^{(L)}+\epsilon)^{\frac{p}{4}-\frac{1}{2}} (\mathbf v_1^{(L)})^*]\nonumber\\
&=[\mathbf h_1^{(1)},\ldots,\mathbf h_1^{(L)}]
\end{align}
Similarly, we can obtain $\mathbf H_2^{\frac{1}{2}}$ as:
\begin{align}\label{eq:h2}
\mathbf H_2^{\frac{1}{2}}&=[(\lambda_{p_2}^{(1)}+\epsilon)^{\frac{p}{4}-\frac{1}{2}} (\mathbf v_2^{(1)})^*,...,(\lambda_{p_2}^{(M)}+\epsilon)^{\frac{p}{4}-\frac{1}{2}} (\mathbf v_2^{(M)})^*]\nonumber\\
&=[\mathbf h_2^{(1)},\ldots,\mathbf h_2^{(M)}]
\end{align}
\subsection{Implementation details}

\begin{figure}[ht!]
\vspace{-0.6em} \centering
\subfigure[SNR of ankle MR image]{\hspace{-0.5em}\includegraphics[width= 0.245\textwidth]{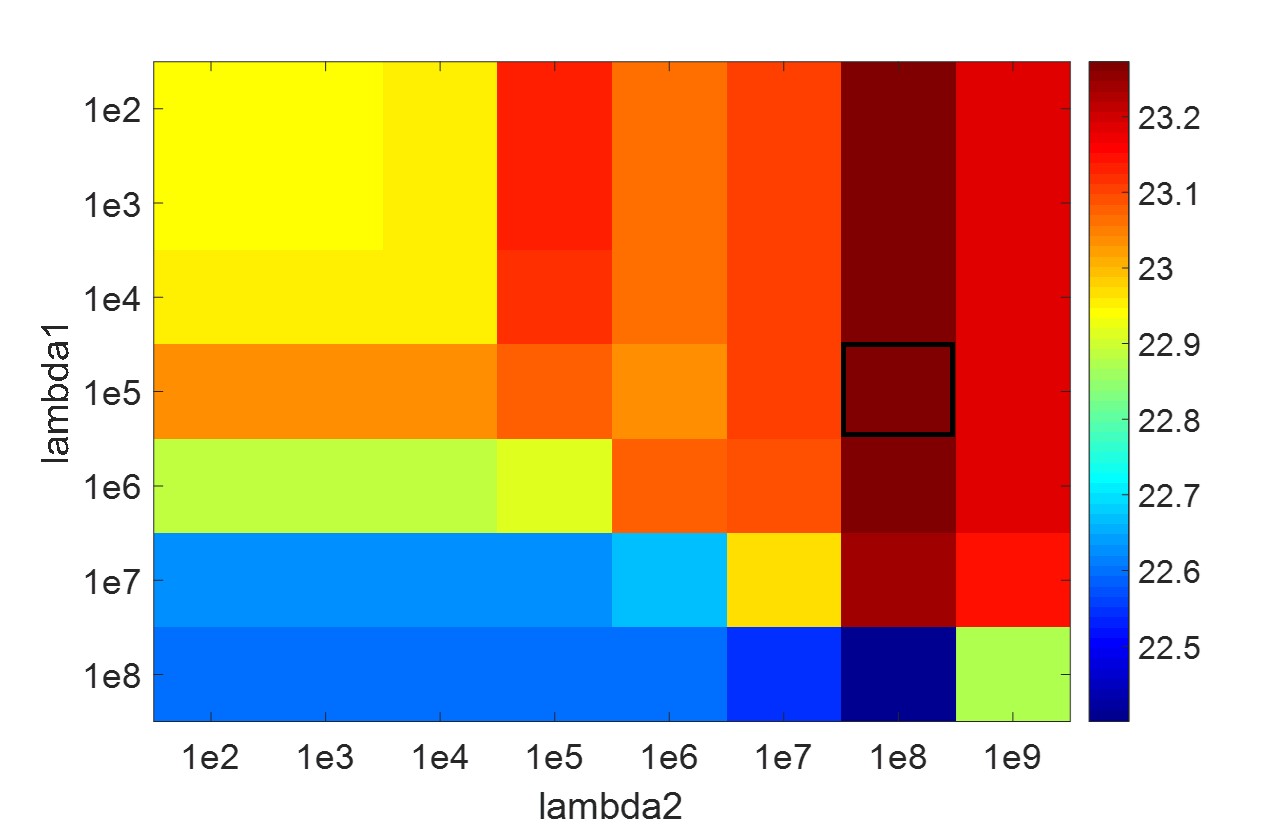}}
\setlength{\fboxrule}{0.5pt}
\subfigure[SNR of phantom image]{\hspace{-0.5em}\includegraphics[width= 0.245\textwidth]{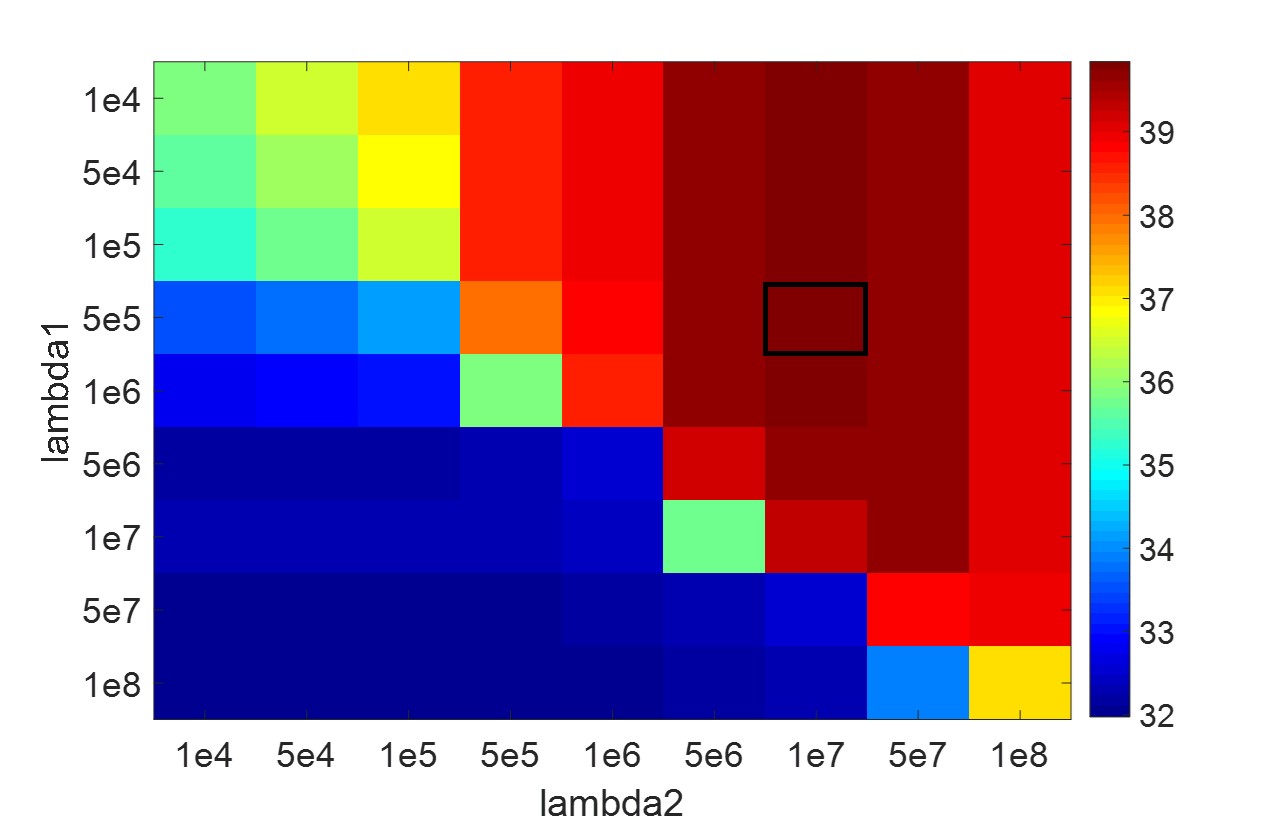}}
\setlength{\fboxrule}{0.5pt}

\caption{Parameters optimization results for the recovery of two images. (a) shows the SNR as the function of the two parameters $\lambda_1$ and $\lambda_2$ for the 6-fold undersampling recovery of an ankle MR image (as shown in Fig. \ref{ankle1}). (b) shows the SNR as the function of $\lambda_1$ and $\lambda_2$ for the 4-fold undersampling recovery of a piecewise smooth phantom image (as shown in Fig. \ref{phantom}). The black rectangles correspond to the optimal parameters with the highest SNR values.  }\label{sequentialfig}\vspace{-0.8em} \label{SNR-para}
\end{figure}

The details for solving the optimization problem (\ref{eq:optprob1}) are shown in the following pseudocode Algorithm 1. In order to investigate how the SNR values of the recovered image behave as the function of the balancing parameters $\lambda_1$ and $\lambda_2$, we plot the parameters optimization results for two images in Fig. \ref{SNR-para}, where (a) correspond to the parameters for the compressed sensing reconstruction of an ankle MR image with the acceleration factor of 6, and (b) correspond to the parameters choices for the recovery of a phantom image with 4-fold undersampling. We find that the tuning of the two parameters is not very time consuming, since we observe that the optimal parameters are localized in a narrow range between $10^5$ and $10^6$ for $\lambda_1$ and between $10^7$ and $10^8$ for $\lambda_2$, for different images under different scenarios.

\begin{algorithm*}
\SetAlgoLined
Initialization: $\hat{\boldsymbol \rho}^{(0)}\leftarrow {\cal A}^*\mathbf b$, $n$ $\leftarrow$ 1, $\epsilon^{(0)}>0$ \;
\While{$n<N_{max}$}{
\textbf{Step1: Update of weight matrices}\;
\Indp
\text{Compute the Gram matrices $\mathbf G_i$ ($i=1,2$)} using (\ref{eq:grammtx1}) and (\ref{eq:grammtx2})\;
\text{Compute eigendecompositions ($\lambda_{p_i}^{(k)},\mathbf v_i^{(k)})_{k=1}^{K}$ of }\
\text{$\mathbf G_i$ ($i=1,2$)}\;
\text{Compute $\mathbf h_1^{(l)}$ and $\mathbf h_2^{(m)}$ using (\ref{eq:h1}) and (\ref{eq:h2})}\;

\Indm
\textbf{Step2: Update of least squares}\;
\Indp
\text{Compute the weight matrices $\mathbf S_1$ and $\mathbf S_2$ using (\ref{eq:s1}) and (\ref{eq:s2})}\;
\text{Solve the following least squares problem :}\
\text{
$\min\limits_{\hat{\boldsymbol\rho}_1,\hat{\boldsymbol\rho}_2}
\lambda_1\|\mathbf S_1^{\frac{1}{2}}\mathbf y_1\|^2_F+\lambda_2\|\mathbf S_2^{\frac{1}{2}}\mathbf y_2\|^2_F
+\|{\cal A}(\hat{\boldsymbol\rho}_1+\hat{\boldsymbol\rho}_2)-\mathbf b \|^2$}\text{using ADMM iterations (\ref{y1}) to (\ref{eq:qi})}\;
\Indm
\text{Choose $\epsilon^{(n)}$ such that $0<\epsilon^{(n)}\leq \epsilon^{(n-1)}$}\;
}
\caption{GSLR image recovery algorithm}
\end{algorithm*}

\section{Experiments and Results}
\label{sec:typestyle}

\subsection{1-D signal recovery}
\begin{figure}[ht!]
\vspace{-0.6em} \centering
\subfigure[Original signal]{\hspace{-0.5em}\includegraphics[width= 0.15\textwidth]{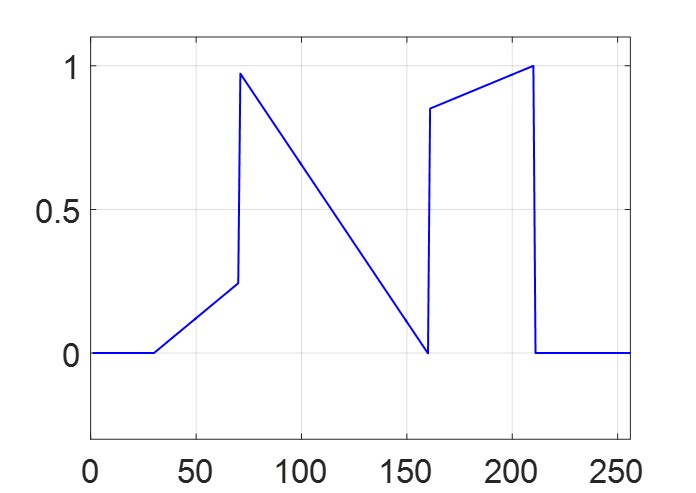}}
\setlength{\fboxrule}{0.5pt}\\
\subfigure[2-fold variable density sampling]{\hspace{-0.5em}\includegraphics[width= 0.15\textwidth,height = 0.1\textwidth]{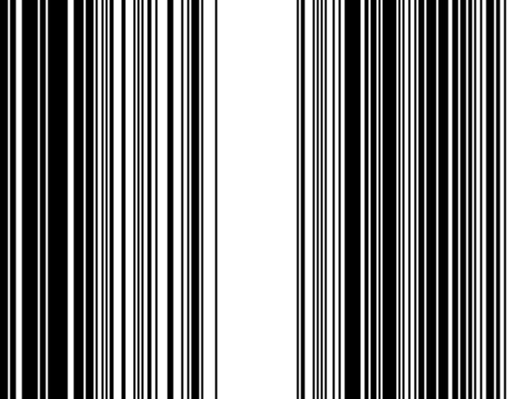}}
\setlength{\fboxrule}{0.5pt}
\subfigure[IFFT signal]{\hspace{-0.5em}\includegraphics[width= 0.15\textwidth]{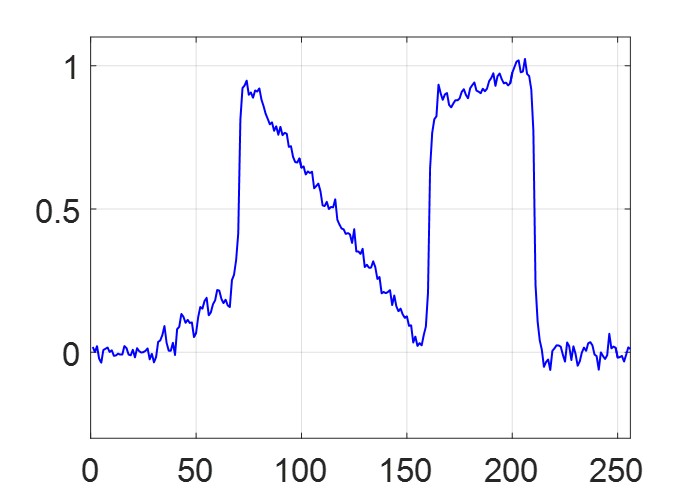}}
\setlength{\fboxrule}{0.5pt}
\subfigure[Reconstruction from (c)]{\hspace{-0.5em}\includegraphics[width= 0.15\textwidth]{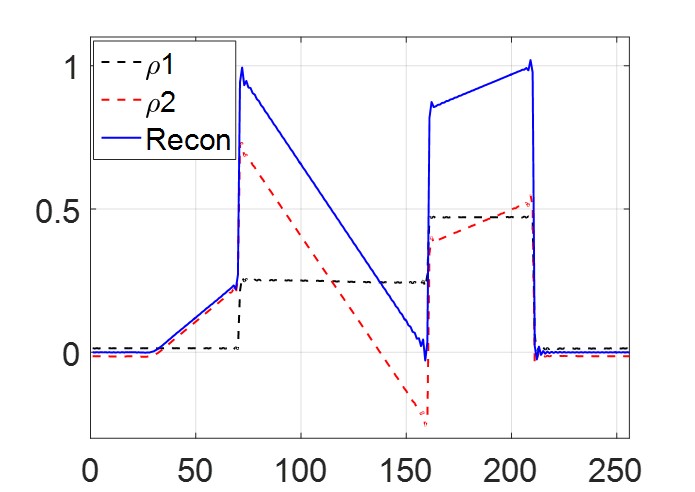}}
\setlength{\fboxrule}{0.5pt}
\subfigure[4-fold random sampling]{\hspace{-0.5em}\includegraphics[width= 0.15\textwidth,height = 0.1\textwidth]{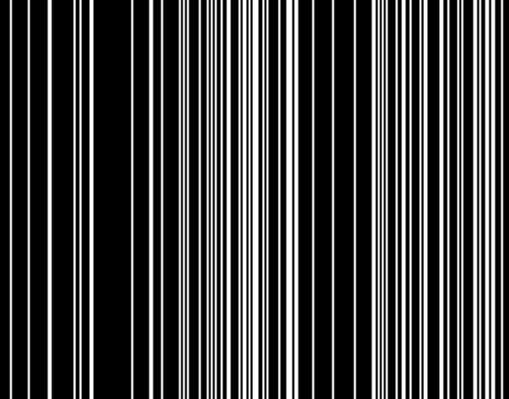}}
\setlength{\fboxrule}{0.5pt}
\subfigure[IFFT signal]{\hspace{-0.5em}\includegraphics[width= 0.15\textwidth]{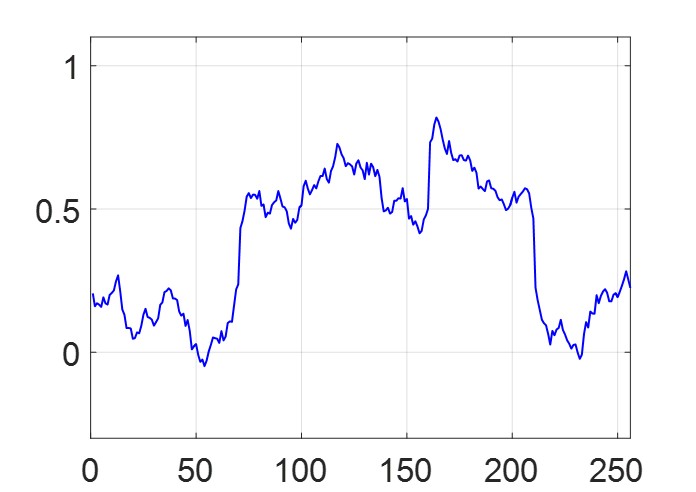}}
\setlength{\fboxrule}{0.5pt}
\subfigure[Reconstruction from (f)]{\hspace{-0.5em}\includegraphics[width= 0.15\textwidth]{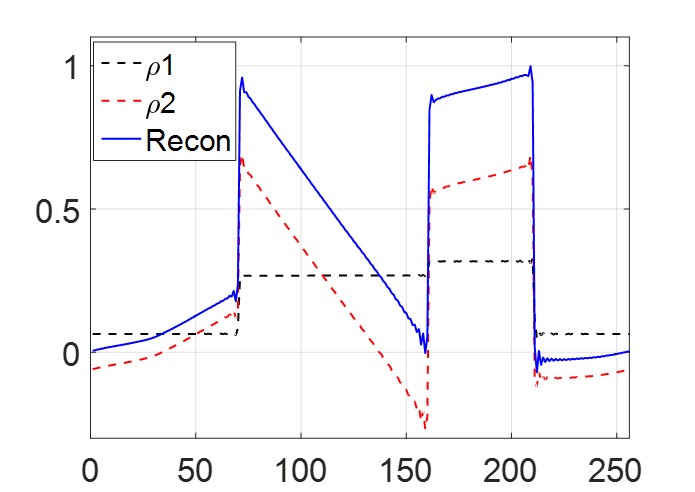}}
\setlength{\fboxrule}{0.5pt}
\caption{1-D signal recovery using the GSLR method. (a) is the original signal, which is undersampled by a 2-fold variable density random undersampling pattern, indicated in (b). (c) is the direct IFFT recovery of the undersampled measurements. (d) shows the recovered signal (in blue solid line) and the decomposition results of the two components, namely the piecewise constant component (showed in black dotted line) and the piecewise linear component (showed in red dotted line). (e)-(g) represent the results of a 4-fold random undersampling compressed sensing signal recovery experiment. Note that the proposed scheme recovers the Fourier coefficients of the signal; the ringing at the edges is associated by the inverse Fourier transform of the recovered Fourier coefficients. }\label{sequentialfig}\vspace{-0.8em} \label{1D-recon}
\end{figure}

We first experiment on a 1-D signal to investigate the performance of the algorithm on recovering signals from their undersampled measurements. Fig. \ref{1D-recon} (a) shows the original signal, which is 2-fold undersampled in $k$ space using variable density undersampling pattern, indicated in (b). The direct IFFT recovery is shown in (c). (d) shows the recovered signal (in blue solid line) using the proposed GSLR method in 1-D, and the decomposition results of the piecewise constant component $\boldsymbol\rho_1$ (in black dotted line) and the piecewise linear component $\boldsymbol\rho_2$ (in red dotted line). We then experiment on the signal recovery using a 4-fold random undersampling pattern, shown in (e). (f) is the direct IFFT of the undersampled measurements. (g) represents the recovered signal and the decomposition results. The results clearly show that by the GSLR method, both the jump discontinuities and the linear parts of the signal are nicely restored.

\subsection{MR images recovery}
The performance of the proposed method is investigated in the context of compressed sensing MR images reconstruction. We compare the proposed GSLR method with the first order and the second order structured low-rank algorithms. We also study the improvement of the image quality offered by the GSLR algorithm over standard TV, TGV algorithm \cite{knoll2011second}, and the LORAKS method \cite{haldar2014low}.
For all of the experiments, we have manually tuned the parameters to ensure the optimal performance in each scenario. Specifically, we determine the parameters to obtain the optimized signal-to-noise ratio (SNR) to ensure fair comparisons between different methods. The SNR of the recovered image is computed as:
\begin{equation}
\mbox{SNR}=-10\log_{10}{\left(\frac{\|f_{\mbox{orig}}-\hat f \|^2_F}{\|f_{\mbox{orig}}\|^2_F}\right)}
\end{equation}
where $\hat f$ is the recovered image, $f_{\mbox{orig}}$ is the original image, and $\|\cdot\|_F$ is the Frobenius norm.
\begin{figure*}[ht!]
\vspace{-0.6em} \centering
{\hspace{-0.3em}\includegraphics[width= 20cm]{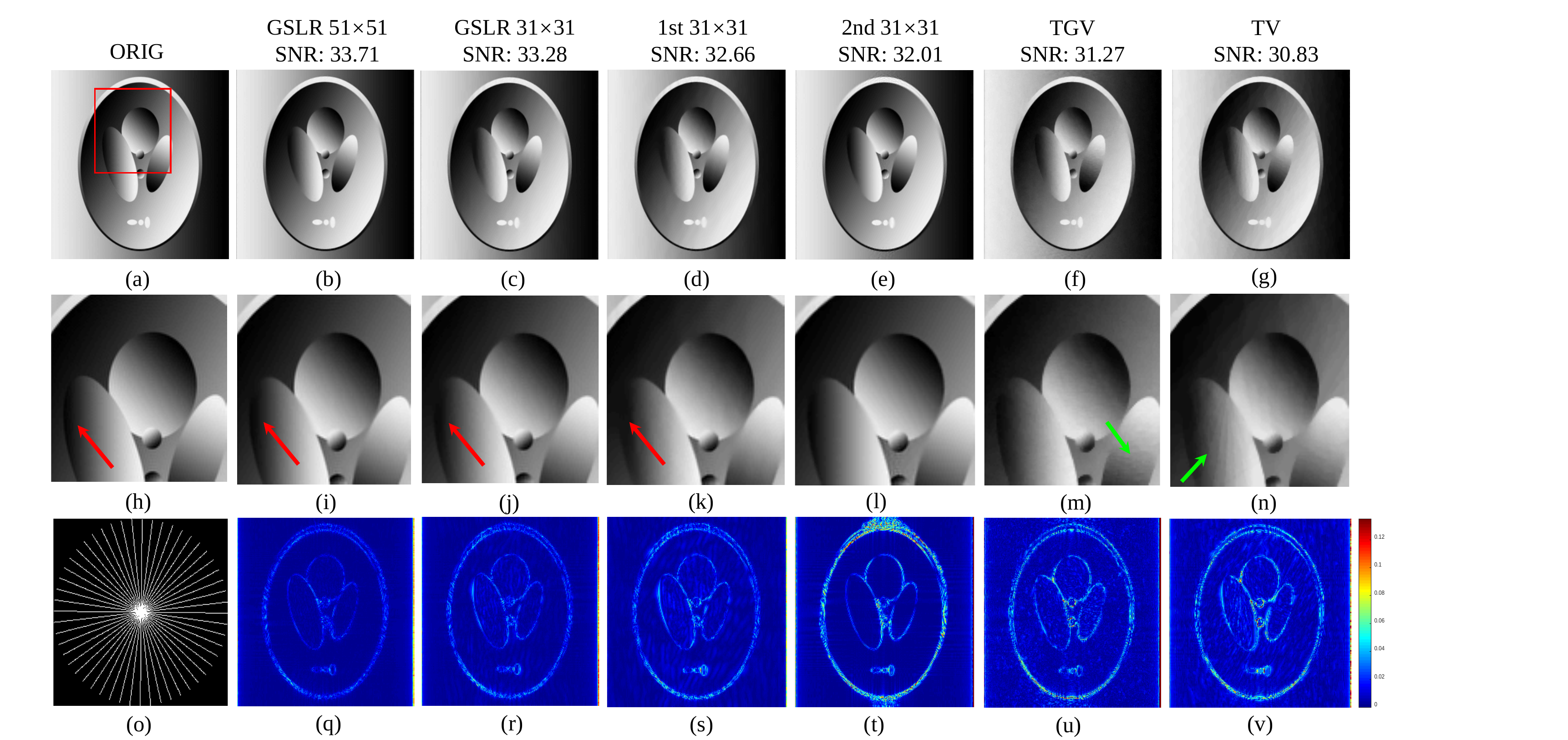}}
\setlength{\fboxrule}{0.5pt}
\vspace{-1.2em}
\caption{Recovery of a piecewise smooth phantom image from around 10-fold undersampled measurements. (a): the actual image. (b)-(g): Reconstructions using the proposed GSLR method with filter size of $51\times 51$, GSLR with filter size of $31\times 31$, the 1st and 2nd SLA with filter size of $31\times 31$, TGV, and the standard TV, respectively. (h)-(n): the zoomed regions of the area indicated in the red rectangle for the corresponding images. (o): the undersampling pattern. (q)-(v): the error images. Note that recovered images using TGV and TV methods present undersampling artifacts, indicated in green arrows, while the GSLR methods outperform the other methods in recovering both the edges and the smooth regions, indicated in red arrows.}
\label{sequentialfig}\vspace{-0.8em} \label{phantom}
\end{figure*}

In the experiments, we consider two types of undersampling trajectory: a radial trajectory with uniform angular spacing, and a 3-D variable density random retrospective undersampling trajectory. For the 3-D sampling pattern, since the readout direction is orthogonal to the image plane, such undersampling patterns can be implemented on the scanner.

We first study the performance of the proposed method for the recovery of a piecewise smooth phantom image from its noiseless $k$ space data.
We assume the data was sampled with 26 $k$ radial spokes, with the approximate acceleration factor of 10.7.
Fig. \ref{phantom} (a) is the actual image, (b) to (g) are the recovered images using GSLR with filter size of $51\times 51$, GSLR with filter size of $31 \times 31$, the 1st SLA and the 2nd SLA with filter size of $31\times 31$, TGV, and the standard TV, respectively. The second row show the zoomed regions of the corresponding images. (o) is the undersampling pattern, (q) to (v) are the error images.

We observe that the structured low-rank algorithms outperform TGV and the standard TV algorithms under this scenario, in that the recovered images by TGV and TV methods suffer from obvious undersampling artifacts, indicated in green arrows. For the structured low-rank algorithms with filter size of $31\times 31$, it is shown that GSLR performs better than the 1st SLA and the 2nd SLA in recovering the edges, indicated in red arrows. It is shown that with larger filter sizes ($51\times 51$), the GSLR method provides the best reconstruction result with the SNR improvement of around 3dB over standard TV.

\begin{figure*}[ht!]
\vspace{-0.6em} \centering
{\hspace{-0.3em}\includegraphics[width= 18cm]{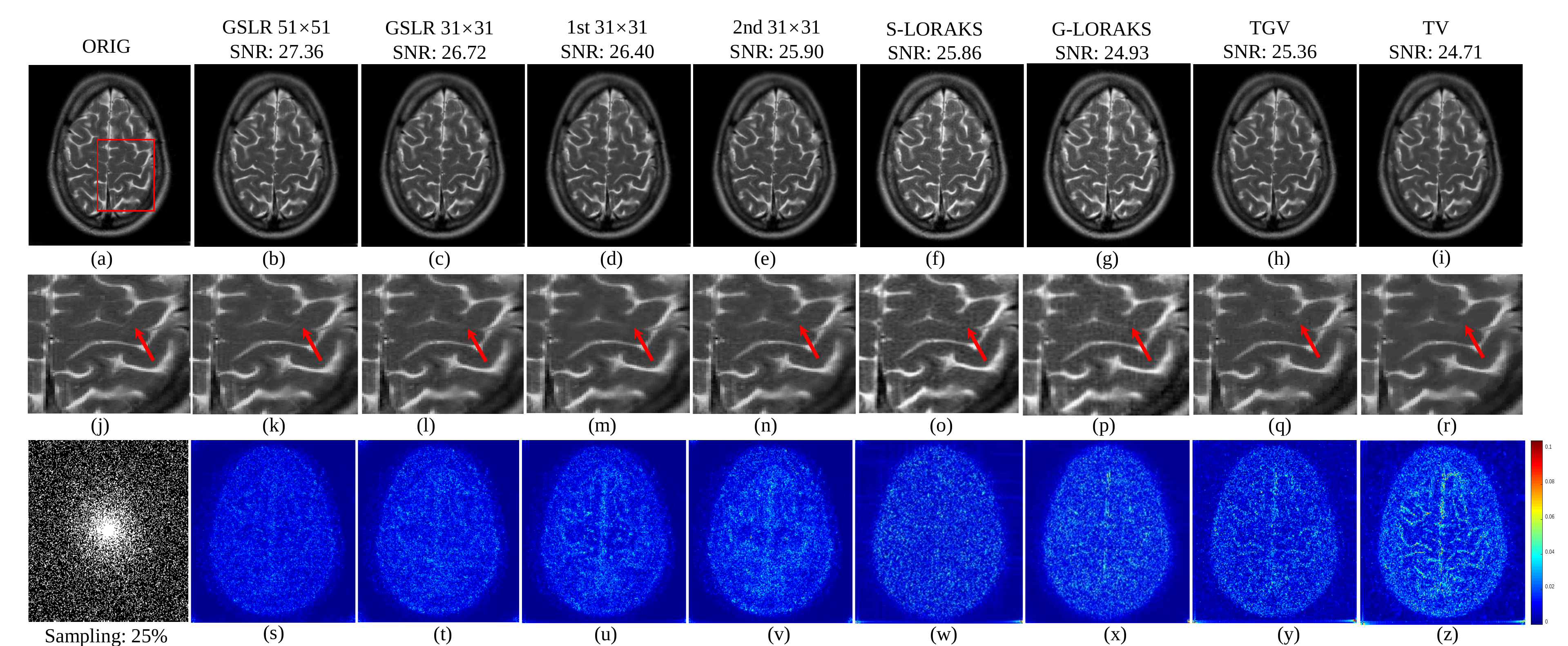}}
\setlength{\fboxrule}{0.5pt}
\vspace{-0.3em}
\caption{Recovery of a brain MR image from 4-fold undersampled measurements. (a): the original image. (b)-(h): Reconstructions using GSLR with filter size of $51\times 51$, GSLR with filter size $31\times 31$, the 1st and 2nd SLA with filter size $31\times 31$, S-LORAKS, TGV, and the standard TV, respectively. (i)-(p): The zoomed versions of the area shown in the red rectangle in (a). (q): The undersampling pattern. (r)-(x): the error images.}
\label{sequentialfig}\vspace{-1.3em} \label{brain1}
\end{figure*}

\begin{figure*}[ht!]
\vspace{0.9em} \centering
{\hspace{-0.3em}\includegraphics[width= 19cm]{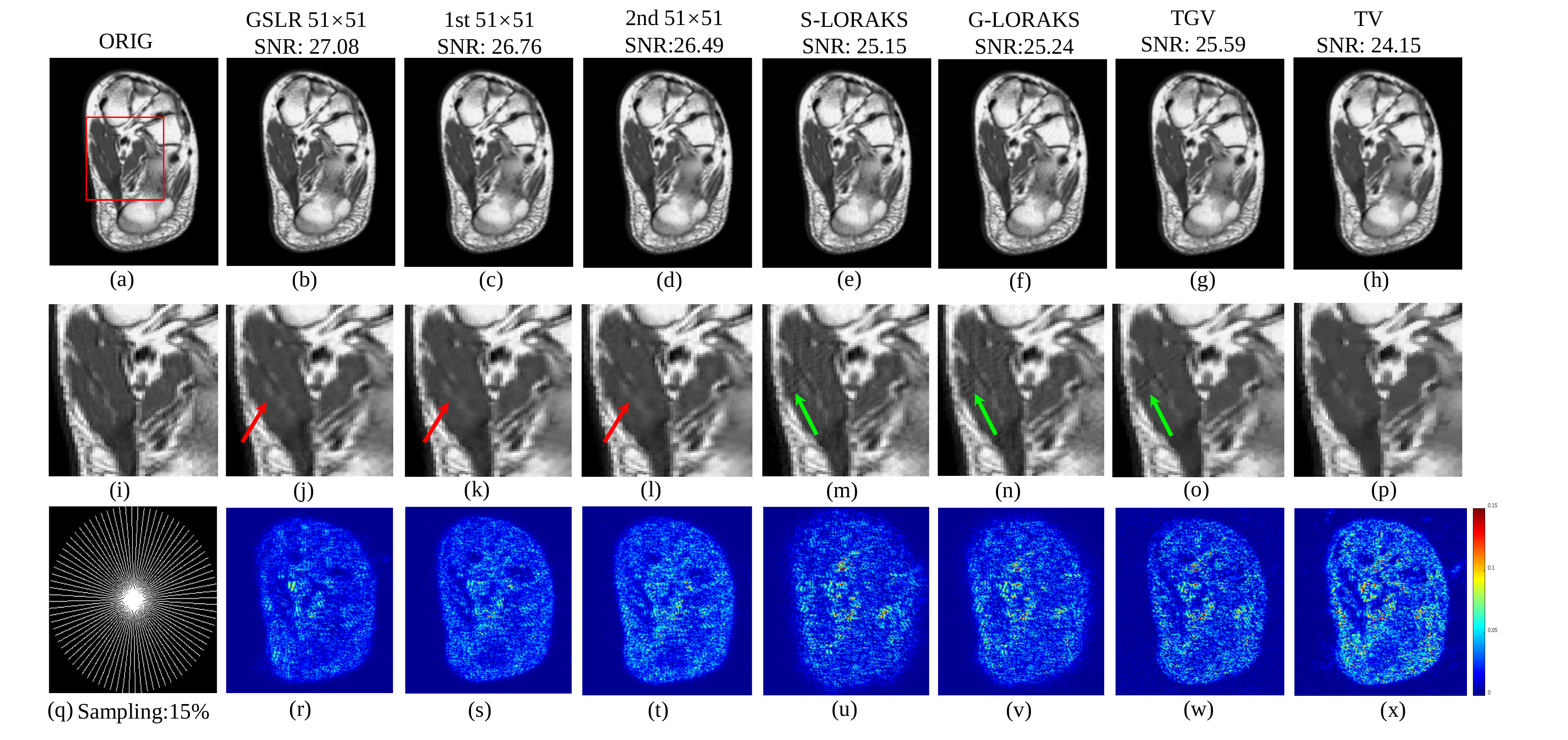}}
\setlength{\fboxrule}{0.5pt}
\vspace{-0.8em}
\caption{Recovery of an ankle MR image from 6.7-fold undersampled measurements with radial undersampling pattern. (a): The actual image. (b)-(f): The recovery images using GSLR, 1st SLA, 2nd SLA with filter size of $51\times 51$, S-LORAKS, G-LORAKS, TGV, and TV, respectively. (i)-(p): The zoomed versions of the images. (q): The undersampling pattern. (r)-(s): Error images using the corresponding methods. Note that the proposed GSLR method performs the best in preserving the edges and eliminating the artifacts caused by the undersampling, compared with the other methods.} \label{sequentialfig}\vspace{-0.8em} \label{ankle1}
\vspace{-0.5em}
\end{figure*}

In the following experiments, we investigate the proposed GSLR method on the reconstruction of single-coil real MR images. The reconstructions of a brain MR image at the acceleration of 4 is shown in Fig.\ref{brain1}, where we compare the proposed GSLR method using different filter sizes with the first and second order SLA, S-LORAKS method, TGV, and the standard TV method. Fig. \ref{brain1} (a) is the original image.
(b) to (h) are the reconstructions using the GSLR with filter size of $51\times 51$, GSLR with filter size $31\times 31$, the 1st and 2nd SLA with filter size $31\times 31$, S-LORAKS, TGV, and the standard TV. (i) to (p) are the zoomed versions of the images, indicated by the red rectangle. (q) is the variable density random undersampling pattern. (r) to (x) are the error images using the corresponding methods. It is seen that among all of the methods, GSLR performs the best in preserving the details and providing the most accurately recovered image. Note that by increasing the filter size from $31\times 31$ to $51\times 51$, the image quality is significantly improved, with only a modest increase in runtime (85 s versus 36 s).

We demonstrate the performance of the proposed method on the reconstruction of an ankle MR image at the approximate acceleration rate of 6.7 using the radial undersampling pattern in Fig. \ref{ankle1}. In this experiment, we compare the proposed GSLR method with 1st and 2nd SLR, S-LORAKS, G-LORAKS, TGV, and the standard TV. All of the structured low rank methods are with filter size of $51\times 51$. (a) is the original image. (b) to (h) are the reconstructed images using GSLR, 1st SLA, 2nd SLA, S-LORAKS, G-LORAKS, TGV, and TV, respectively. (j) to (p) are the zoomed versions of the images indicated by the red rectangle shown in (a). (q) is the radial undersampling pattern. (r) to (x) are the error images. We observe that TV method gives blurry reconstruction. The images recovered by S-LORAKS, G-LORAKS, and TGV methods have undersampling artifacts, indicated in the green arrow. The structured low rank methods provide improved results, among which GSLR performs better in preserving image details, shown in red arrows.

\begin{figure*}[ht!]
\vspace{-0.2em} \centering
{\hspace{-0.3em}\includegraphics[width= 18cm]{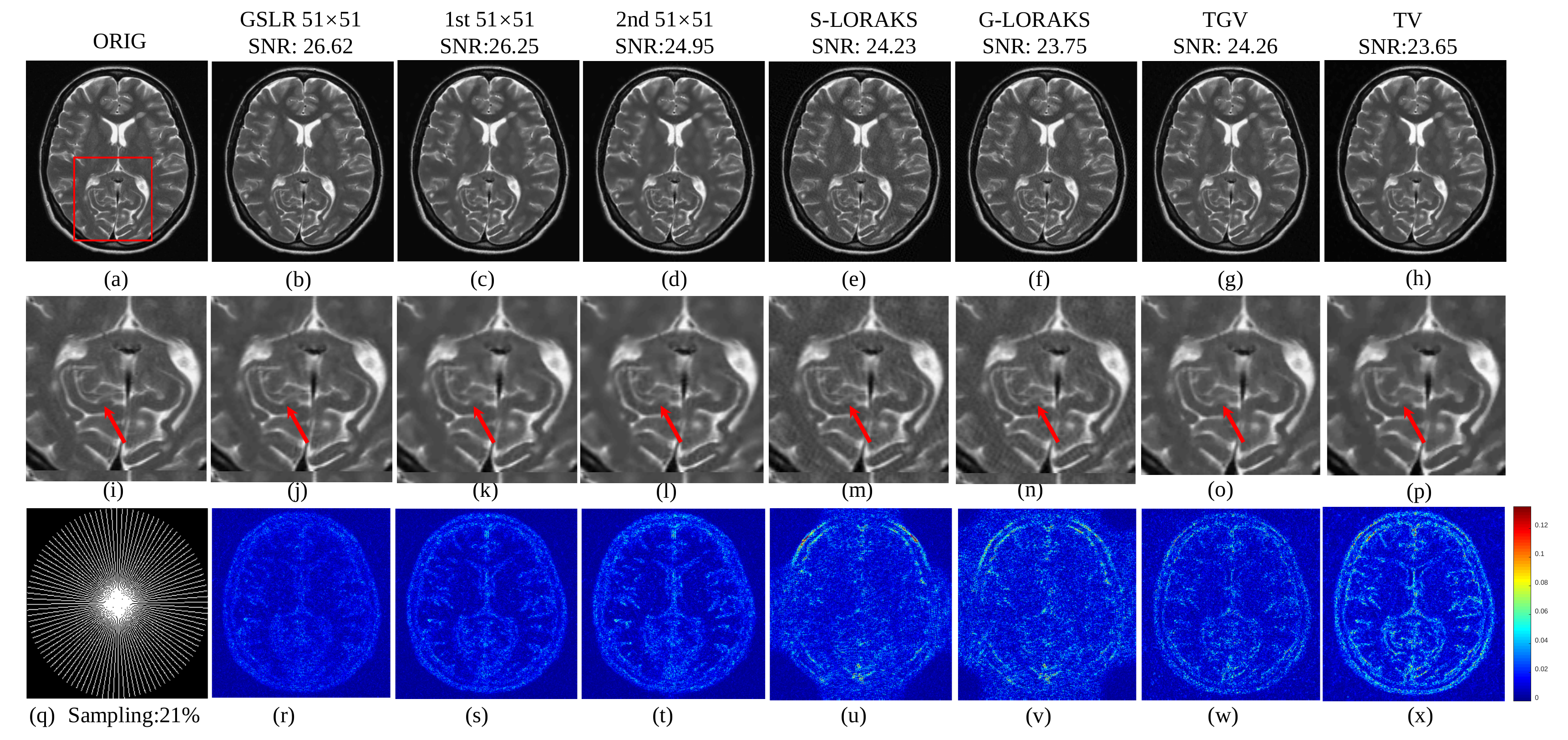}}
\setlength{\fboxrule}{0.5pt}
\caption{Recovery of a brain MR image from 4.85-fold undersampled measurements. (a): The actual image. (b)-(h): The recovery images using GSLR, 1st and 2nd SLA with $51\times 51$ filter size, S-LORAKS, G-LORAKS, TGV, and the standard TV, respectively. (i)-(p): Zoomed versions of the red rectangular area for different methods. (q): the undersampling pattern. (r)-(x): Error images of the corresponding methods.} \label{sequentialfig}\vspace{-0.8em} \label{brain2}
\end{figure*}

In Fig. \ref{brain2}, we experiment on a brain MR image using radial undersampling pattern with the acceleration factor around 4.8. (a) is the actual image. (b) to (h) are the reconstruction results using GSLR with filter size of $51\times 51$, 1st and 2nd SLA with filter size of $51\times 51$, S-LORAKS, G-LORAKS, TGV, and the standard TV, respectively. The second row are the zoomed versions of the red rectangular area shown in (a) for different methods. (q) shows the undersampling pattern. (r) to (x) are the error images of the corresponding methods, respectively. We observe that TV and TGV methods provide blurry reconstructions, and the LORAKS methods preserve fine details better while suffers from undersampling artifacts. Among the methods, GLSR provides the best reconstruction and improves the SNR by around 2dB compared over standard TV.

\begin{figure*}[ht!]
\vspace{-0.6em} \centering
{\hspace{-0.3em}\includegraphics[width= 18.5cm]{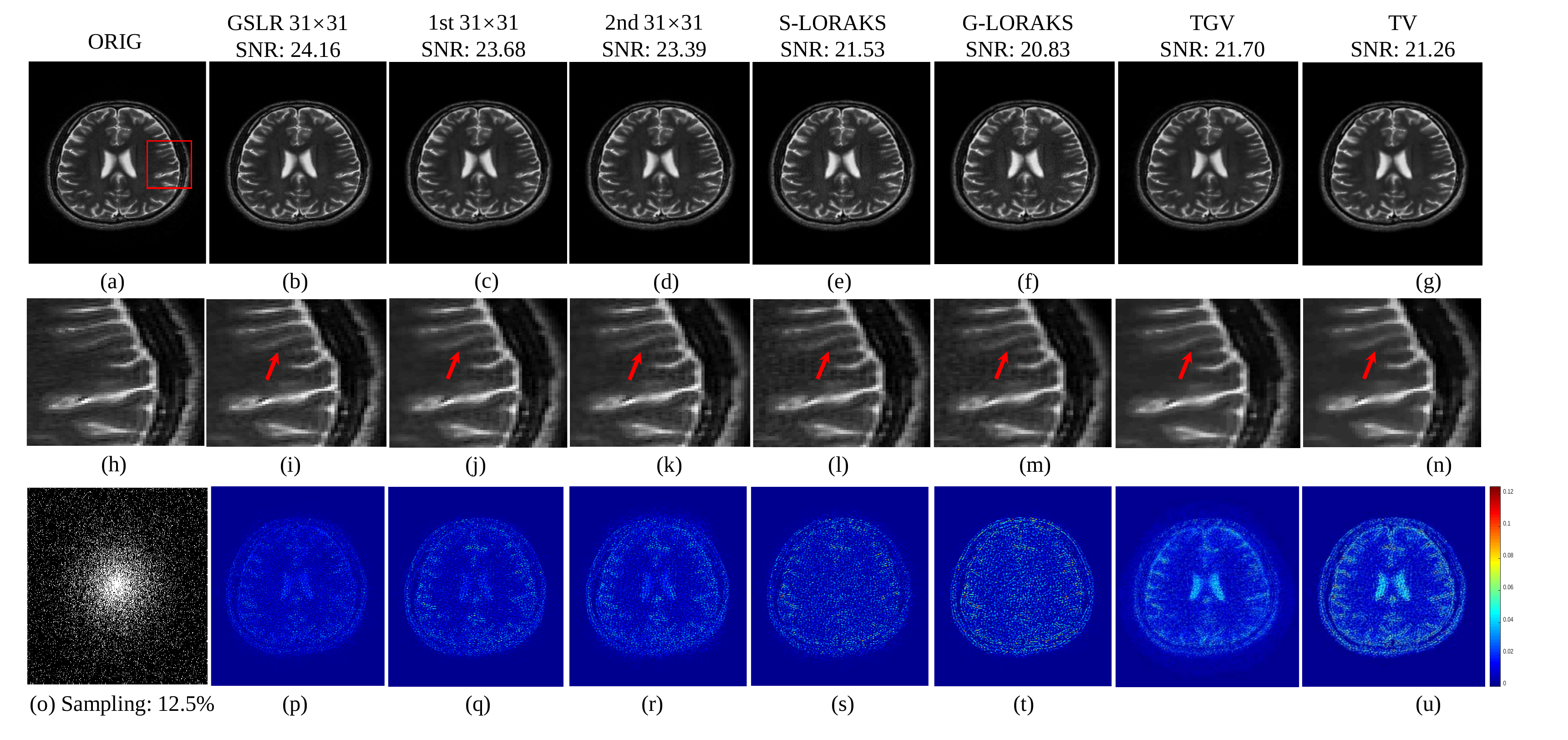}}
\setlength{\fboxrule}{1.5pt}
\vspace{-1.5em}
\caption{Recovery of a multi-coil brain MR dataset from 8-fold undersampled measurements. (a): The actual image. (b)-(g): The reconstructions using GSLR, the first and second order SLA with $31\times 31$ filter size, S-LORAKS and G-LORAKS method, and the standard TV. (h)-(n): The zoomed regions. (o): The undersampling pattern. (p)-(u): The error images. Note that GLSR provides the most accurate reconstruction result compared with the other methods, indicated by red arrows.}\label{sequentialfig}\vspace{-0.1em} \label{multi}\vspace{-1.5em}
\end{figure*}

\begin{figure}[ht!]
\vspace{-0.2em} \centering
{\hspace{-0.3em}\includegraphics[width= 9cm]{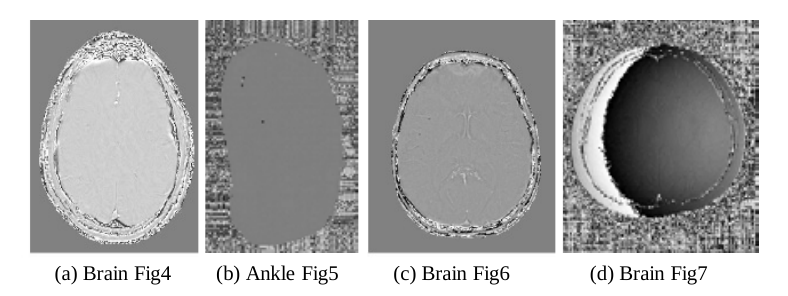}}
\setlength{\fboxrule}{0.5pt}
\vspace{-0.95em}
\caption{Phase images of the real datasets. (a) Brain image in Fig.4. (b) Ankle image in Fig.5. (c) Brain image in Fig.6. (d) Brain image in Fig.7.} \label{sequentialfig}\vspace{-0.8em} \label{phase}
\end{figure}

In Fig. \ref{multi}, we compare different methods on the recovery of a multi-coil MR dataset acquired using four coils from 8-fold undersampled measurements. The data was retrospectively undersampled using the variable density random undersampling pattern.
(a) is the actual image. (b) to (g) show the reconstruction results using GSLR, 1st and 2nd SLA with filter size of $31\times 31$, S-LORAKS, G-LORAKS, and the standard TV, respectively. (h) to (n) are the zoomed regions indicated in the red rectangle. (o) is the undersampling pattern. (p) to (u) indicate error images by different methods.
According to the results, we observe that compared with the other methods, GLSR performs the best in preserving the image features and providing the  recovered image with highest SNR value.
We have shown the phase images of all the datasets in Fig. \ref{phase}. We note that all of the images are associated with reasonable phase variations, expected from a typical MR acquisition. We note that GSLR relies on the compact representation of the image, enabled by its decomposition into piecewise constant and linear components. Since S-LORAKS and G-LORAKS do not exploit these property, we obtain improved reconstructions with filter sizes larger than 31x31.

The SNRs of the recovered images using variable density random undersampling patterns are shown in Table \ref{SNRresults}, and the reconstruction results using radial undersampling patterns are shown in Table \ref{SNRresults_radial}. We compare the 1st and 2nd SLA, S-LORAKS, G-LORAKS, TGV, the standard TV with the proposed GSLR method. For the structured low-rank algorithms, we compare the performance by different filter sizes. Specifically, we use three different filter sizes, $15\times 15$, $31\times 31$ and $51\times 51$ for the variable density undersampling experiments, and two filter sizes for the radial undersampling experiments. Note that when the filter size is $15\times 15$, the results provided by GSLR are not comparable to the other methods for some cases. However, using larger filter sizes leads to significantly improved image quality. For filter sizes $31\times 31$ and $51\times 51$, GSLR consistently obtains the best results, with the SNR improvement by around 2-3 dB over standard TV.
The reason is that the size of the filter specifies the type of curves or edges that its zero set can capture. Specifically, smaller filters can only represent simpler and smoother curves, while larger filters can represent complex shapes (see \cite{ongie2016off} for an illustration). When complex structures are presented in the image, the use of a smaller filter fails to capture the intricate details. We note that for most images, we need to use larger filters to ensure that the details are well captured. The use of larger filters is made possible by the proposed IRLS algorithm, which does not require us to explicitly compute the Toeplitz matrices.



\begin{table*}[ht!] \footnotesize
\centering
\normalmarginpar
\marginnote{R2}[0cm]
\begin{tabular}{c||c|c|c||c|c|c||c|c|c}
\hline\hline
{\bf filter size} &    {\bf [15,15]} &   {\bf [31,31]}  & {\bf [51,51]}&   {\bf [15,15]} &   {\bf [31,31]} & {\bf [51,51]}&  {\bf [15,15]} &   {\bf [31,31]} &{\bf [51,51]}\\
\hline
     {\bf Phantom} &      \multicolumn{ 3}{c||}{{\bf acc=2}} & \multicolumn{ 3}{c||}{{\bf acc=4 }}& \multicolumn{ 3}{c}{{\bf acc=5 }} \\
\hline
  {\bf First SLA} &      47.63 &      53.78 &  55.34&    38.06 &       44.20&  46.20 &    34.87 &       39.93 & 42.06 \\

{\bf Second SLA} &      48.77 &      56.26 &57.66 &      33.73 &      43.30&  45.53&     33.62 &       38.07& 41.10 \\

{\bf TGV} &    42.17   &    42.17   & 42.17&  37.86     &   37.86   &37.86 &      35.40 &  35.40&    35.40 \\

{\bf TV} &      41.31 &      41.31 &  41.31&    35.88 & 35.88&     35.88&      35.06 & 35.06&      35.06   \\


{\bf GSLR} &  {\bf 51.13} & {\bf 56.94} &{\bf58.21}&  {\bf39.22} & {\bf 45.31}& {\bf47.18}& {\bf35.93} & {\bf 40.79} &{\bf43.11}\\

%
%
%

\hline
    {\bf Brain Fig.\ref{brain1} } &      \multicolumn{ 3}{c||}{{\bf acc=2}} & \multicolumn{ 3}{c||}{{\bf acc=4 }}& \multicolumn{ 3}{c}{{\bf acc=5 }}\\
\hline
  {\bf First SLA} &      34.93 &       36.14 & 36.72&     24.85 &       26.40&26.80&      23.00 &       24.66&25.24\\

{\bf Second SLA} &      34.10 &      35.67 & 36.31&     24.28 &      25.90& 26.43&     22.63 &       24.30&24.73 \\

{\bf TGV} &     33.09       &   33.09  &33.09  &      25.36 &      25.36 &  25.36&    23.34 & 23.34&      23.34\\
{\bf TV} &  32.40  &   32.40&     32.40 &      24.71 & 24.71&   24.71 &  22.90 &22.90&     22.90\\

{\bf S-LORAKS} & {\bf35.43} &35.43 & 35.43&{\bf 25.86} & 25.86& 25.86& 24.22 & 24.22 &24.22\\
{\bf G-LORAKS} & 33.36 & 33.36 &33.36& 24.93 & 24.93& 24.93& 23.01 &23.01 &23.01\\

{\bf GSLR} &  35.34 &      {\bf 36.38}&{\bf 37.19} &  25.34  &  {\bf26.72}  &{\bf 27.36} &  {\bf23.58}  &  {\bf25.03}&{\bf 26.07}\\

\hline
    {\bf Brain Fig.\ref{brain2} } &      \multicolumn{ 3}{c||}{{\bf acc=2}}&      \multicolumn{ 3}{c||}{{\bf acc=4}} & \multicolumn{ 3}{c}{{\bf acc=5 }}\\
\hline
  {\bf First SLA} &      30.31 &       32.30&  32.61&    20.64 &       23.63 & 24.17&     18.77 &       21.36&22.54\\

{\bf Second SLA} &      30.16 &       31.99&32.33&      19.24 &      22.81 & 23.62&     17.56 &      20.70&22.19 \\

{\bf TGV} &   30.26    &   30.26    & 30.26&      {\bf23.05} &      23.05 & 23.05&     {\bf21.13} &      21.13&21.13 \\

{\bf TV} &      30.10 & 30.10&      30.10&       22.65 &22.65&      22.65 &      20.83 &  20.83&    20.83 \\

{\bf S-LORAKS} & 29.83 &29.83 &29.83&22.71 & 22.71& 22.71& 21.02 & 21.02 &21.02\\
{\bf G-LORAKS} & 28.69 & 28.69 &28.69& 22.06 & 22.06& 22.06& 20.69 &20.69 &20.69\\

{\bf GSLR} & {\bf 30.70} &      {\bf 32.51}& {\bf33.17}& 21.41 &      {\bf 24.20} & {\bf24.93}& 19.51  &  {\bf22.15}&{\bf 23.08}\\

\hline
    {\bf Ankle  } &      \multicolumn{ 3}{c||}{{\bf acc=2}}&      \multicolumn{ 3}{c||}{{\bf acc=4}} & \multicolumn{ 3}{c}{{\bf acc=5 }}\\
\hline
  {\bf First SLA} &      37.80 &       38.13&38.42 &      30.01 &       30.89 &31.17&      27.37 &       28.26 &28.50\\

{\bf Second SLA} &      37.96 &       38.38 & 38.61&     29.65 &      30.69 &30.90&      26.65 &       27.31&28.08 \\

{\bf TGV} &      36.89 &      36.89&36.89&      30.43 & 30.43&     30.43 &      {\bf28.05} &28.05&       28.05\\

{\bf TV} &      33.43 & 33.43&      33.43 &      28.38 & 28.38&     28.38 &      26.22 & 26.22&      26.22 \\
{\bf S-LORAKS} & 37.91 &37.91 &37.91&30.18 & 30.18& 30.18& 27.44 & 27.44 &27.44\\
{\bf G-LORAKS} & 37.02 & 37.02 &37.02& 29.27 & 29.27& 29.27& 26.85 &26.85 &?26.85\\
{\bf GSLR} & {\bf 38.05} &      {\bf 38.47} &{\bf39.05}&  {\bf 30.46} &      {\bf 31.00} &{\bf31.66}& 27.96 &  {\bf 28.43}&{\bf28.96}\\

\hline
 {\bf Multi-coil } & \multicolumn{ 3}{c||}{{\bf acc=4 }} & \multicolumn{ 3}{c||}{{\bf acc=6 }} & \multicolumn{ 3}{c}{{\bf acc=8 }}\\
\hline
  {\bf First SLA} &      29.64 &      30.97& 31.32&     23.48 &      25.45 &25.81&       21.43 &       23.68&24.04\\

{\bf Second SLA} &      29.70 &      31.22&31.68&      23.74 &      25.67&25.99&      21.02 &       23.39&23.62\\

{\bf TGV} &      27.48 & 27.48&     27.48&      21.53 & 21.53&     21.53 &      21.70 & 21.70&      21.70\\

{\bf TV} &      26.68 & 26.68&     26.68&      22.15 & 22.15&     22.15 &      21.26 & 21.26&      21.26\\

{\bf S-LORAK} &      27.83 & 27.83&     27.83&      {\bf23.92} &      23.92 &23.92&   21.53    &21.53&   21.53    \\

{\bf G-LORAKS} & 27.22 & 27.22 &27.22& 22.81 & 22.81& 22.81& 20.83 &20.83 &20.83\\
{\bf GSLR} & {\bf 30.08} & {\bf 31.48} &{\bf 32.24}&  23.88 & {\bf 25.82} &{\bf26.29}& {\bf 22.00}& {\bf 24.16}&{\bf 24.58}\\

\hline \hline
\end{tabular}  \caption{Comparison of MR image recovery algorithms using variable density random undersampling pattern}
\vspace{-1.5em}
\label{SNRresults}
\end{table*}

\begin{table}[ht!] \footnotesize
\centering
\normalmarginpar
\marginnote{R2}[0cm]
\begin{tabular}{c||c|c||c|c}
\hline\hline
{\bf filter size} &   {\bf [31,31]} & {\bf [51,51]}&  {\bf [31,31]} &{\bf [51,51]}\\

\hline
    {\bf Brain Fig.\ref{brain1} } & \multicolumn{ 2}{c||}{{\bf acc$\approx$4.8 }}& \multicolumn{ 2}{c}{{\bf acc$\approx$6.7 }}\\
\hline
  {\bf First SLA} &26.64&     26.85 &  23.19&23.80\\

{\bf Second SLA} & 26.05&     26.55 &   22.82&23.32 \\

{\bf TGV}   &  25.90&    25.90 & 23.11&  23.11\\
{\bf TV}  &   25.26 &  25.26 &22.59&     22.59\\

{\bf S-LORAKS} & 26.31& 26.31 & 22.85 &22.85\\
{\bf G-LORAKS}& 25.97& 25.97 &21.30 &21.30\\

{\bf GSLR}    &{\bf 26.77} &  {\bf27.25}  & {\bf23.45}&{\bf 24.18}\\

\hline
    {\bf Brain Fig.\ref{brain2} } &      \multicolumn{ 2}{c||}{{\bf acc$\approx$4.8}} & \multicolumn{ 2}{c}{{\bf acc$\approx$6.7 }}\\
\hline
  {\bf First SLA} & 24.99&     26.25 &  21.17&22.62\\

{\bf Second SLA} & 23.82&  24.95 &  20.86&22.29 \\

{\bf TGV}  & 24.26&    24.26 &      21.63&21.63 \\

{\bf TV}&      23.65 &      23.65 &  20.88&    20.88 \\

{\bf S-LORAKS}& 24.23& 24.23 & 21.20 &21.20\\
{\bf G-LORAKS}& 23.75& 23.75 &21.53 &21.53\\

{\bf GSLR} & {\bf25.23}& {\bf26.62}  &  {\bf22.47}& {\bf23.01}\\

\hline
    {\bf Ankle  } &      \multicolumn{ 2}{c||}{{\bf acc$\approx$4.8}} & \multicolumn{ 2}{c}{{\bf acc$\approx$6.7 }}\\
\hline
  {\bf First SLA} &31.02&  31.37 &  26.15 &26.76\\

{\bf Second SLA}  &31.14&  31.42 &  26.07&26.49 \\

{\bf TGV} &  30.53 &   30.53 &25.59&  25.59\\

{\bf TV} &  28.53 &   28.53 & 24.15&   24.15 \\
{\bf S-LORAKS}& 31.03& 31.03 & 25.15 &25.15\\
{\bf G-LORAKS}& 29.89& 29.89 &25.24 &25.24\\
{\bf GSLR}&{\bf 31.39}& {\bf31.69} &  {\bf 26.60}&{\bf 27.08}\\

\hline
 {\bf Multi-coil } & \multicolumn{ 2}{c||}{{\bf acc=5.2 }} & \multicolumn{ 2}{c}{{\bf acc=10 }}\\
\hline
  {\bf First SLA}  &27.78& 29.01 &  22.33&22.87\\

{\bf Second SLA} &27.24&  28.15 &  21.76&22.60\\

{\bf TGV}&   26.08&    26.08 & 20.77&   20.77\\
{\bf TV}&   24.92&    24.92 & 18.53&   18.53\\

{\bf S-LORAK} &27.32 & 27.32   &21.01& 21.01  \\

{\bf G-LORAKS}& 26.58& 26.58 &20.22 &20.22\\
{\bf GSLR} &{\bf 27.90}  & {\bf 29.43}& {\bf22.51} &{\bf23.15}\\

\hline \hline
\end{tabular}  \caption{Comparison of MR image recovery algorithms using radial undersampling pattern}
\label{SNRresults_radial}
\vspace{-1.5em}
\end{table}






\section{Conclusion}
\label{sec:majhead}

We proposed a novel generalized structured low-rank algorithm to recover images from their undersampled $k$ space measurements. We assume that an image can be modeled as the superposition of two piecewise smooth functions, namely a piecewise constant component, and a piecewise linear component. Each component can be annihilated by multiplication with a bandlimited polynomial function, which yields to a structured Toeplitz matrix. We formulate a combined regularized optimization algorithm by exploiting the low-rank property of the Toeplitz matrix. In order to solve the corresponding problem efficiently, we adapt the iteratively reweighted least squares method which alternates between the computation of the annihilation filter and the least squares problem. We investigate the proposed algorithm on the compressed sensing reconstruction of single-coil and multi-coil MR images. Experiments show that the proposed algorithm provides more accurate recovery results compared with the state-of-the-art approaches.

\bibliographystyle{IEEEtran}
\bibliography{acl}

\end{document}